\documentclass[prd,twocolumn,superscriptaddress,nofootinbib,showpacs]{revtex4}

\usepackage{graphicx,amsmath,amsfonts,amssymb,revsymb,dcolumn,epsfig,bm}

\begin{document}

\title{Depletion of energy from Naked Singular regions during
gravitational collapse}

\author{S. Jhingan}\email{sanjay.jhingan@gmail.com}
\affiliation{Centre for Theoretical Physics, Jamia Millia Islamia,
New Delhi 110025, India}

\author{I. H. Dwivedi}\email{dwivedi@gmail.com}
\affiliation{21, Ballabhji Vihar, Dayalbagh, Agra, India}

\author{Sukratu Barve}\email{sukratu@cms.unipune.ac.in}
\affiliation{Centre for Modeling and Simulation, University of Pune,
Pune 411007, India}

\begin{abstract}
A distinguishable and observable physical property of Naked Singular
Regions of the spacetime formed during a gravitational collapse has
important implications for both experimental and theoretical
relativity. We examine here whether energy can escape physically
from naked singular regions to reach either a local or a distant
observer within the framework of general relativity. We find that in
case of imploding null dust collapse scenarios field outgoing singular null geodesics
including the cauchy horizon can be immersed between two Vaidya spacetimes
as null boundary layers with non vanishing positive energy density.
Thus energy can transported from the naked singularity to either a local or a distant observer. And example illustrating that similar considerations
can be applied to dust models
is given.\end{abstract}


\maketitle

\section{Introduction}
A star with sufficient remnant mass ($ \gtrsim 3 M_{\odot}$), on
completion of its nuclear fuel cycle, must enter the phase of a
continuous gravitational collapse. Once the nuclear fuel is
exhausted gravitational forces become all powerful and hence star's
internal pressure can not sustain the equilibrium resulting in a
continued collapse \cite{dutt,oppen}. In the late stages of collapse
the gravitational forces become dominant and the physics of collapse
is determined mainly by the theory of general relativity. Under
quite general and physical situations general relativity predicts
that such a collapse must end in a singularity, i.e., a region of
spacetime with extreme curvatures \cite{pen65,HW66,HW}. Physically
one could describe singularity as a region of space with vanishing
volume and unbounded gravitational forces. General relativity,
however, does not say anything about the nature or physical
properties of such a singularity. This is partially due to the fact
that mathematical structure breaks down preventing analysis at and
beyond the singularity. One could perhaps argue that as collapse
progresses and matter is condensed in a region comparable to Planck
length the quantum physical properties of spacetime would become
dominant, thus preventing the formation of singularity. But this
picture may not hold also since gravity as a force is very different
in its nature in comparison to other forces and has a geometrical
interpretation as curvature of spacetime. Moreover, despite numerous
efforts, a viable quantum theory of gravity is not in sight. Hence
for such regions of spacetime, whether relativity theory or quantum
physics would determine the physics is still an open question.

To fill in the gap in our understanding of spacetime singularities
in a mathematical consistent manner, a cosmic censorship
conjecture, that all gravitational collapse must end in a
black hole was proposed \cite{pen65,Pen}. The physical consequence of
such a hypothesis is that even before the formation of a singularity
a trapped surface develops covering the singularity from the outside
world. Hence from a physical point of view singularity is hidden
from the outside world. Initial studies in censorship were
directed towards formulating the conjecture in a mathematically precise
manner which could then possibly be proven \cite{wald_rev}. This also led to
formulation of other conjectures like, hoop conjecture by Kip Thorne
and Siefert's conjecture \cite{kip,seifert}. However, extensive
studies in collapse with various forms of matter fields have shown
that under fairly generic reasonable physical conditions both naked
singularity and black holes would form as an end state of collapse,
depending on various initial and boundary conditions
\cite{JJS}. It
is still not very clear how to classify either matter or the initial
and boundary conditions in a satisfactory way which would end in
either state of singularity (naked or covered). Thus from
the studies this far almost all physically reasonable matter fields lead
to both naked and covered singularities during collapse (see,
\cite{joshi_book}, and references therein).

Considerable work has since been done on naked singularities from
the point of view of giving counterexamples to cosmic censorship but
also on the study of their nature and structure. Having established
their existence it is important to study the phenomena of formation
of naked singularity from a more astrophysical perspective. One
could look for a possible observable signature of naked singularity
distinguishing them from other compact strong gravity objects, like
black holes. In the studies carried out this far the stress has been
towards showing that for a naked singularity to be ``observable'' a
family of lightlike geodesics must terminate at the singularity
\cite{Christo_all,jd}. Optical appearance and redshift for such
possible radiation has also been studied \cite{red-shift}. However,
from the point of view of general relativity the first null ray
coming out of singularity forms a Cauchy horizon (CH), and the
spacetime model cannot remain valid after its formation. Therefore,
without any consistent extension of spacetime beyond CH the validity
and usefulness of all such geodesic analysis becomes doubtful. The
basic question of the existence of the spacetime structure after the
CH is unaddressed (it is difficult to provide extensions of
spacetimes, for example, even for shell-crossing singularities which
are gravitationally weak \cite{clarke_extension}) which is of utmost
importance if we want to talk about families of geodesics ending at
singularity in the past, making it a possible astrophysical source.

In this paper we wish to study the structure of the spacetime from
this perspective.  Is it possible to connect the two spacetimes
before and after with Cauchy horizon as the boundary? Whether the
resulting spacetime after the CH has formed can still have the same
symmetry? Does relativity theory allows such continuation of
spacetime through CH and whether boundary conditions pose any
restrictions? Furthermore, can these boundary layers carry energy
from naked singularity to a distant observer? Earlier Hiscock et
al. has considered a model spacetime in which cauchy horizon ultimately becomes
the event horizon of the schwarzschild black hole with non vanishing surface
energy density and where it could be
visible to observers falling into the blackhole \cite{Hiscock}.

If indeed the formation of a naked singularity is a physical
phenomenon then the CH would represent a null surface layer
emanating from the naked singularity, and reaching the distant
observer separating the two spacetimes. It has been
suggested in various studies that naked singularities may be
responsible for various high energy phenomena in our universe (for
example gamma ray bursts etc. \cite{gammaburst}). It has also been
suggested that in the late stages of collapse, when spacetime
shrinks to size of the order of planck length quantum effects would
play a dominant role resulting in either a burst of particle
creation or preventing the formation of singularity all together
\cite{quantumeffects}. Our aim in this paper is to examine two
examples of naked singularities within the frame work of general
relativity and whether this allows such a scenario as emission of a
impulsive null wave carrying energy from the naked singularity. The
result of such a study would have manifold implications. First
does there exist a spacetime after the formation of a naked
singularity which can be joined satisfactorily together with the
original model separated by the null shockwave (CH)? If
such a spacetime exists then whether it allows the existence of
outgoing families of geodesics terminating at the singularity in the
past. Second, and equally important, question is the structure of
the CH itself. Whether this null surface `boundary layer' is
allowed to carry huge amounts of energy along the null ray to distant
observer? And, if the answer is in affirmative, what is its
structure and whether this scenario can be called a valid solution
to the Einstein equations?

\section{A Collapsing Star}

Despite numerous exact solutions of the  field equations, very few
exact solutions of the field equations exist which can describe a
physically reasonable collapsing matter cloud. In fact, in nearly
all the studies of spherically symmetric collapse, the key models
are either Lema\`itre -Tolman-Bondi metric (LTB) \cite{TBL} or the Vaidya
spacetime \cite{Vaidya}. Both these spacetimes have been well
studied, and very well may be the only physically reasonable exact solution
available. In all such studies it has been shown that there are out
going null and time like geodesics which terminate at the naked
singularity in the past. The visibility of the singularity in terms
of a \emph{roots equation} whose roots are tangents to the out-going
radial null geodesics with past end points at the singularity.
Therefore, problem of relating initial data with end state of
collapse is reduced to finding roots of a polynomial equation
\cite{jd,roots-form}.

We would first take up the Vaidya spacetime.
Existence of naked singularity in this model is well
established \cite{vaidya_visible}. In particular for the case of a
imploding shell with a linear mass function $M(v)=\lambda v,$ for
$\lambda \leq 1/8$ singularity is known to be a naked singularity.

The metric describing a spherically symmetric Vaidya space-time is given by
\begin{equation}\label{eq:vaidya}
ds^2= -\left (1 - \frac{2M(v)}{r} \right) dv^2 + 2dv dr + r^2 d \Omega^2,
\end{equation}
where $d\Omega^2=d\theta^2+\sin^2\theta d\phi^2$. For linear mass
function case $2M(v)=\lambda v$, and the singularity formed at $v=0$, $r=0,$ is
naked iff:
\begin{equation}
x^2-\frac{x}{\lambda}+\frac{2}{\lambda}=0
\end{equation}
has real and positive root where $x=\frac{v}{r}$. It follows that
for $\lambda \le 1/{8,} $ the above has two real and positive roots
namely $( \gamma, \beta)$, $ \beta >\gamma$ given by

\begin{equation}\label{CH0}
\gamma= \frac{1 - \sqrt{1 -
8{\lambda}}}{2\lambda}, \; \beta= \frac{1 + \sqrt{1 -
8{\lambda}}}{2\lambda}  .
\end{equation}
CH is the first null geodesic given by
\begin{equation}\label{CH}
v = \gamma r .
\end{equation}
while a family of geodesics which terminate at the singularity in
the past with the tangent $x=\beta$ are given by

\begin{equation}\label{geo-fam}
r = V\frac{(\beta-x)^{\frac{\gamma}{\beta - \gamma}}
}{(x-\gamma)^{\frac{\beta}{\beta - \gamma}}
} \, .
\end{equation}
Where V is a parameter (constant along out going null geodesics) labeling different
geodesics of the family.

Once the singularity forms the spacetime below the CH is described
by the metric above. However, if the further analysis of family of
geodesics is to be valid than the spacetime beyond CH must also be
described by similar metric with CH as the boundary between the two
solutions. If such a collapse scenario is to be called a solution of
the field equations, the two Vaidya spacetimes separated by the CH
(null hypersurface layer) must form a smooth solution. Hence, the
thin null shell  with the stress energy should be matched with two
spacetimes before and after. Barrab\`es and Israel (and Poisson)
have analyzed in detail the conditions for immersion of such null
surface layers between two general spherically symmetric spacetimes.
To implement our model we follow the prescription of matching across
null hypersurface by Barrab\`es and Israel \cite{israel} (see also
Poisson \cite{poisson}).

\begin{figure}[ht]
\includegraphics[width=7.0cm,angle=0]{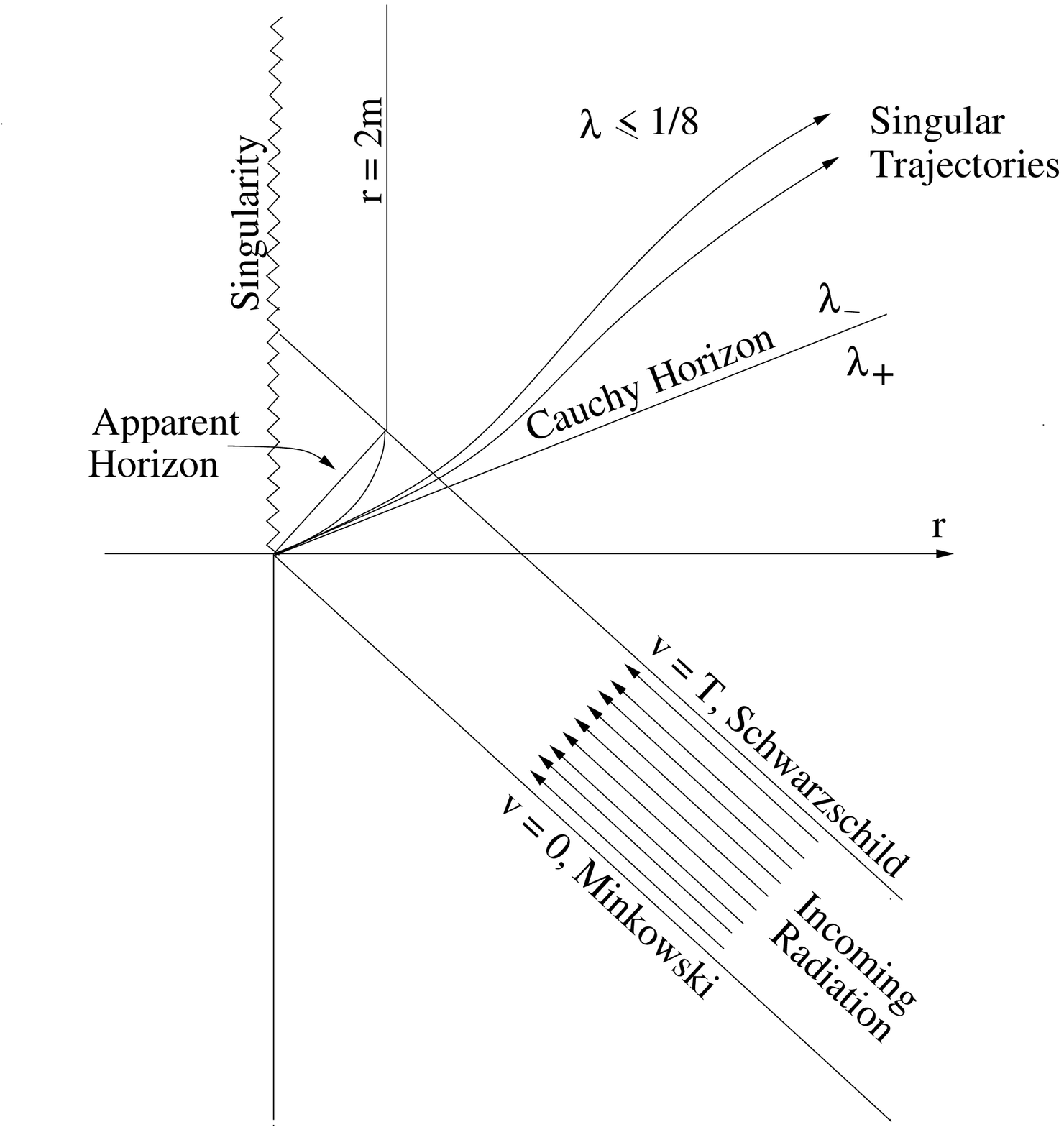}
\caption{Naked singularity forming in the radiation collapse
}\label{fig:vaidya}
\end{figure}

Let the two spacetimes separating the first singular light ray (CH) be given by
$\lambda_+$ before and $\lambda_-$ after (see Fig \ref{fig:vaidya}).
We can describe the spacetime metric across CH in the
following form
\begin{eqnarray}\label{metrics}
ds_I^2 & = & -(1-\lambda_{+}x_+)  du_{+}^2 + 2 du_{+} dr
+r^2 d\Omega^2 \\
ds_{II}^2 & = & -(1-\lambda_{-}x_-) du_-^2 + 2 du_- dr +r^2 d\Omega^2
\end{eqnarray}
where $x_+ = u_+/r$ and $x_- = u_-/r$. Here region I and region II
correspond to spacetime before and after formation of Cauchy
horizon, respectively. In order to glue these two Vaidya spacetimes
along the null hypersurface $\Sigma$ (CH) we should have
\begin{eqnarray}\label{blI}
x_+ = \gamma_+ &=& \frac{1}{2 \lambda_+}\left[1-\sqrt{1-8{\lambda_+}}
\right],\nonumber \\  \frac{d u_+}{dr} &=& \left.\frac{2}{(1-\lambda_+ x_+)}\right|_\Sigma ,
\end{eqnarray}
in spacetime I, and
\begin{eqnarray}\label{blII}
x_- = \gamma_- &=& \frac{1}{2 \lambda_-} \left[ 1 - \sqrt{1 - 8 \lambda_-}
\right] , \nonumber\\ \frac{d u_-}{dr} &=& \left.\frac{2}{(1-\lambda_-
x_-)}\right|_\Sigma
\end{eqnarray}
in spacetime II.
On the boundary we have from continuity
\begin{equation}\label{conti}
u_+|_{\Sigma} = \frac{\gamma_+}{ \gamma_-}u_-|_{\Sigma} ,
\end{equation}

Defining tangent vectors on the Cauchy horizon :
\begin{eqnarray}
k^a\equiv e^a_{(1)}&=& \left.\left[ \frac{2 r}{ (r-\lambda_+ u_+)},
1 , 0 , 0 \right]\right|_\Sigma, \; \mbox{region I} \nonumber \\
k^a\equiv e^a_{(1)}&=& \left.\left[ \frac{2 r}{(r-\lambda_- u_-)}, 1
, 0 , 0 \right]\right|_\Sigma, \; \mbox{region II} \nonumber
\end{eqnarray}
and
\begin{equation}
e^a_{(2)}= \frac{\partial}{\partial \theta}, \qquad e^a_{(3)} =
\frac{\partial}{\partial \phi} \; .
\end{equation}
for region I \& II. Where $ y^\alpha=(r,\theta,\phi)$ are the
intrinsic coordinates on $ \Sigma\; (\alpha =1,2,3), $ and we take $r$
to be the parameter of the null generator. The transverse vectors
completing the basis for region I and II are given by
\begin{eqnarray}\label{trans}
N^a &=& \left.\left[0,-\frac{1}{2}\left(1-\frac{\lambda_+ u_+}{r}\right),0,0 \right]\right|_\Sigma ,
\mbox{region I} \nonumber \\
N^a &=& \left.\left[0,-\frac{1}{2}\left(1-\frac{\lambda_-
u_-}{r}\right),0,0 \right]\right|_\Sigma,  \mbox{region II}
\nonumber \end{eqnarray}
satisfying
\begin{equation}
N_a N^a = 0, \quad N_a k^a = -1, \quad N_a e^a_{(A)}=0,
\end{equation}
and where $(A)=\{\theta,\phi\}$. The transverse curvature
$C^{\pm}_{AB} \; , $ and the intrinsic metric of the surface layer
($\sigma_{AB}$) is given by
\begin{equation}
\sigma_{AB} dx^A dx^B = r^2 (d \theta^2 + \sin^2 \theta d\phi^2 )\; ,
\end{equation}
and
\begin{equation}
C_{AB} =- N_{\alpha} e^{\alpha}_{(A);\beta} e^{\beta}_{(B)}\; .
\end{equation}
We find the surface energy density and pressure of the null layer for Vaidya
case as
\begin{eqnarray}
\mu &=& \sigma^{AB} C_{AB}= \frac{[M]}{4 \pi r^2}=
\frac{\lambda_-u_- - \lambda_{+} u_+}{4 \pi r^2}\;, \nonumber \\
p & =& [h(\lambda,M)] \;.
\end{eqnarray}
First we note that $h(\lambda, M)$ quantifies jump in pressure across
the CH. Since it is transverse component to Cauchy
surface it does not affect the physics of energy propagating along
the CH which is of interest to us here. If energy is
transported along the CH the energy density $\mu$, of
the null layer must be positive definite. It follows from the
continuity of the boundary layer from Eqs. (\ref{blI}) and (\ref{blII}) that
\[
\frac{[M]}{4 \pi r^2}= \frac{u_+}{\gamma_+}\left[\frac{\sqrt{1-8\lambda_+} - \sqrt{1-8\lambda_-}}{8 \pi r^2} \right]\;.
\]
Hence, CH can carry energy to either a local or a distant observer.
Therefore, as a result the rate of collapse slows
down ($\lambda_-<\lambda_+$), which results in a net positive energy
density on the CH. Furthermore, this surface energy on
the CH has a clear physical interpretation. To see this
consider the motion of a freely falling timelike observer (four
velocity $u^a, u^au_a=-1$) in Vaidya spacetime given in Eq. (\ref{eq:vaidya}) (analysis of timelike trajectories in Vaidya spacetime has been worked out
\cite{roots-form}).

\begin{eqnarray}
u^a&\equiv& e^a_{(1)}= \left[ \frac{P}{ r}, \frac{(1-\lambda
x)P}{2r} -\frac{r}{2P} , 0 , 0 \right], \; \mbox{region I} \nonumber
\\ P &=& {(c-s)\pm \sqrt{(c-s)^2+r^2x(2+\lambda x^2-x) }\over
(2+\lambda x^2-x)}, \; \mbox{region II} \nonumber
\end{eqnarray}

Where $c$ is a constant labeling different timelike geodesics and s
is the affine parameter. Positive sign solutions
terminate at the singularity $r=u=0$ with a positive definite tangent $x=\beta$
and hence do not intersect the CH . For all timelike radial observers intersecting the CH we have

\begin{equation}
u_ak^a={r\over (1-\lambda x)P} \; ,
\end{equation}

and therefore at the cauchy horizon we have

\begin{equation}
[u_ak^a]=[u_ae^a_{(A)}]=0 \;.
\end{equation}

It follows that in imploding null dust collapse CH can be immersed between two
Vaidya spacetimes (with linear mass
function) with the parameter $\lambda_+\le \lambda_-$. In case  when $ \lambda_+= \lambda_{-}$
the matching across $\Sigma$ is smooth and no energy is carried along the first ray. In the case otherwise
the rate of collapse slows down and facilitates the positive surface density on the null boundary.

 It has
been shown that there is a family of out going geodesics which
terminate in the past at the singularity with a definite tangent
$x=\beta > \gamma$. The path of such outgoing null geodesics has
been calculated earlier (see \cite{vaidya_visible}) and is
given by Eq. (\ref{geo-fam}).
The problem can be considerably simplified if we can write the
metric in terms of out-going null geodesics. In this representation
CH corresponds to a constant value of one of the coordinates.   Let
us consider a general spherically symmetric spacetime $M^{\pm}$
given by
\begin{equation}
ds^2= -e^{2\Psi} \left (1 - \frac{2m}{r} \right) d V^2 + 2 \zeta
 e^{\Psi} dV dr + r^2 d\Omega^2.
\end{equation}
Here $\psi_{\pm}, m_{\pm}$ are functions of $V_{\pm}$ and $r$. Null
layers given by $V_{\pm}= constant$ are out going if $\zeta=-1$ and
ingoing if $\zeta = 1$. The density and pressure of the null shell
surface immersed in the two spacetimes is
\begin{eqnarray}
\mu &=& \sigma^{AB} C_{AB}= -\zeta \frac{[m]}{4 \pi r^2} , \nonumber \\
p & =& -\zeta \frac{1}{8\pi} \frac{\partial \psi}{\partial r} .
\end{eqnarray}

In order to analyse the case of family of null geodesics let us
consider a coordinate transformation for the spacetime given in
equation (1) $v \rightarrow V , r \rightarrow r $. The Vaidya metric the
for spacetime $M_{\pm}$ now becomes
\begin{equation}
ds^2= -e^{\psi}\left[ e^{\psi} \left(1 - \frac{2m(V,r)}{r} \right)
dV^2 + 2 dV dr \right] + r^2 d \Omega^2,\nonumber
\end{equation}
with metric function $\exp(\psi)$
\begin{equation}\label{metricf}
e^{\psi_{\pm}} = \frac{r\lambda_{\pm}(\beta_{\pm} - x_{\pm})(x_{\pm} - \gamma_{\pm})}{V_{\pm}
(1 - \lambda_{\pm} x_{\pm})} \; ,
\end{equation}
where $2m_{\pm} (V_{\pm},r) = \lambda_{\pm} v_{\pm} (V_{\pm},r)$ and
$\psi_{\pm} = \psi_{\pm} (V_{\pm},r)$. Here $ V_{\pm} = \text{constant}$ are
outgoing singular geodesics with normal $k^a=\delta^a_r$. Hence
outgoing singular null layers immersed between the two Vaidya
spacetimes with different mass functions ($\lambda_+ < \lambda_- <
\frac{1}{8}$) have non vanishing surface density $\mu$ and $p$
allowing energy to escape. Though for lightlike shells there is no
rest frame and therefore $\mu$ and P cannot be given an absolute
meaning as surface density and pressure nonetheless as rightly
pointed out by Israel they serve perfectly well to determine the
results of measurement by any observer In this regard as shown by
Israel that for a radially freely falling observer momentum normal
to the shell is continuous and the energy density associated with
the shell as measured by this radially freely falling observer
($u^a={dx^a}/{d\tau} = [{\dot u}, {\dot r},0,0]) $ is given by
\begin{equation}
T^{ab}_{\Sigma} \;u_a u_b = \frac{[m]}{ 4\pi r^2} \delta (\tau)(k^a u_a)
\end{equation}
and is accompanied by equal energy flux. Here $\tau=0$ is the equation of
$\Sigma$.

We briefly consider now another scenario, namely the inhomogeneous dust
collapse. The metric describing a spherically symmetric space-time
is given by
\begin{equation}\label{ltbmetric}
ds^2= -dt^2 + \frac{R'^2}{1+f}dr^2 + R^2 d \Omega^2,
\end{equation}
where $d\Omega^2 = d \theta^2+\sin^2\theta d\phi^2$, and $R=R(t,r)$ and
$f=f(t,r)$ are arbitrary functions of $t$ and $r$. The metric in
(\ref{ltbmetric}) has to satisfy field equations which can be put in
a form
\begin{equation}
\dot R =  - \sqrt{f + \frac{F}{R}}
\end{equation}
The functions, $F=F(r)$, $R=R(t,r)$ and $f(r) > -1$ are $C^2$
functions throughout the cloud. Notation (') and $(^.)$ are used to
denote partial differentiation with respect to $r$ and $t$. Consider
marginally bound case $f(r)=0$, and $F(r)=F_0 r$. Existence of naked
singularity in this case ($\beta < 3$) is well established (see
\cite{joshi_book} and ref. therein). Function $F = F(r)$ is
interpreted as the mass function and for physical reasons $F(r)\ge
0$, $F'(r)\ge 0$, and gives mass enclosed in a given shell of
co-moving radius $r$. The CH is a null ray $ R = x_0 r$
where $x=x_0$ is lowest of the real and positive root of the algebraic equation

\begin{equation}
2x^4 + x^3 \sqrt{F_0} - 2x + 2 \sqrt{F_0} = 0
\end{equation}

The two dust spacetimes $M_{\pm}$ separating the first singular
light ray (CH) be given by respective mass functions, $F_{-}(r) =
F_0r$ before, and $F_+(r)  = r P^2(r)$ after, where $P(r)$ satisfies
\begin{equation}
P(a+bP)^{c}= \text F_1\; r \;,
\end{equation}
where $F_1$ is a constant and  coefficients can be determined as  $a = 2 x_0^{3/2}$, $b = {(x_0^{3/2}+2)}/{(x_0^{3/2}-1)}$ and
$c = -3 x_0^{3/2}/(x_0^{3/2}+2)$.

An argument similar to one for Vaidya model shows that that cauchy horizon can be immersed between these two two different dust solutions with a non vanishing surface energy density given by different values of constants $F_1$
and $F_0$.

\end{document}